\documentclass[10pt,preprint]{emulateapj}

\def\msun{\rm M_{\sun}}

\def\lsun{\rm L_{\sun}}
\def\micron{$\mu$m}

\def\mdot{\rm \dot{M}}

\begin{document}

\shortauthors{Espaillat et al.}
\shorttitle{UX Tau A \& LkCa 15}

\title{On the Diversity of the Taurus Transitional Disks: UX Tau A \& Lk Ca 15}

\author{C. Espaillat\altaffilmark{1}, N. Calvet\altaffilmark{1}, P.
D'Alessio\altaffilmark{2}, J. Hern\'{a}ndez\altaffilmark{1,3}, C.
Qi\altaffilmark{4}, L. Hartmann\altaffilmark{1}, E. Furlan\altaffilmark{5,6}, \&
D. M. Watson\altaffilmark{7}}

\altaffiltext{1}{Department of Astronomy, University of Michigan, 830 Dennison
Building, 500 Church Street, Ann Arbor, MI 48109, USA; ccespa@umich.edu, ncalvet@umich.edu, hernandj@umich.edu, lhartm@umich.edu}
\altaffiltext{2}{Centro de Radioastronom\'{i}a y Astrof\'{i}sica, Universidad
Nacional Aut\'{o}noma de M\'{e}xico, 58089 Morelia, Michoac\'{a}n, M\'{e}xico; p.dalessio@astrosmo.unam.mx}
\altaffiltext{3}{Centro de Investigaciones de Astronomia, Merida 5101$-$A,
Venezuela}
\altaffiltext{4}{Harvard-Smithsonian Center for Astrophysics, 60 Garden Street,
Cambridge, MA 02138, USA; cqi@cfa.harvard.edu}
\altaffiltext{5}{NASA Astrobiology Institute, and Department of Physics and
Astronomy, UCLA, Los Angeles, CA 90095, USA; furlan@astro.ucla.edu}
\altaffiltext{6}{NASA Postdoctoral Fellow}
\altaffiltext{7}{Department of Physics and Astronomy, University of Rochester,
NY 14627-0171, USA; dmw@astro.pas.rochester.edu}

\begin{abstract}

The recently recognized class of ``transitional disk" systems consists
of young stars
with optically-thick outer disks but inner disks which are mostly
devoid of small dust.
Here we introduce a further class of ``pre-transitional disks" with significant
near-infrared excesses which indicate the presence of an optically thick
inner disk separated from an optically thick outer disk; thus,
the spectral energy distributions of pre-transitional disks suggest the
incipient
development
of disk gaps rather than inner holes.
In UX Tau A, our analysis of the {\it Spitzer} IRS spectrum finds that the
near-infrared excess is produced by an inner optically thick disk and a gap of
$\sim$56 AU is present.
The {\it Spitzer} IRS spectrum of LkCa 15 is suggestive of a gap of $\sim$46 AU,
confirming previous millimeter imaging.  
In addition, UX Tau A contains crystalline silicates in its disk
at
radii $\gtrsim$ 56 AU which
poses a challenge to our understanding of the production of this
crystalline material. 
In contrast, LkCa 15's silicates are amorphous and pristine.  
UX Tau A and LkCa 15 increase our knowledge of the diversity of dust clearing 
in low-mass star formation.

\end{abstract}

\keywords{accretion disks, stars: circumstellar matter, stars: formation, stars:
pre-main sequence}

\section{Introduction}

Previous studies have revealed stars with 
inner disks that are
mostly devoid of small dust, and these ``transitional disks" have been proposed as the 
bridge 
between Class II objects, young stars surrounded by
full disks accreting material onto the central star, and Class III objects,
stars where the protoplanetary disk is mostly dissipated and accretion has
stopped (e.g. Strom et al. 1989, Skrutskie et al. 1990; Stassun et al. 2001).  


New spectra 
from the {\it Spitzer Space
Telescope} which greatly improve our resolution in the infrared have 
been used to define the class of ``transitional disks" as those
with spectral energy distributions (SEDs) characterized by 
a significant deficit of flux in the near-infrared relative to optically
thick full disks, and a substantial infrared
excess in the mid- and far-infrared.
Extensive modeling studies of several transitional disks around T Tauri
stars \citep{dalessio05, uchida04, calvet05, espaillat07} and F-G stars
\citep{brown07} have been presented.
In particular, the SEDs of
the transitional disks of the T Tauri stars (TTS) CoKu
Tau$/$4 \citep{dalessio05}, TW Hya \citep{calvet02, uchida04}, GM Aur, DM Tau
\citep{calvet05}, and CS Cha \citep{espaillat07} have been explained by modeling
the transitional disks with truncated optically thick disks with most of the
mid-infrared emission originating in the inner edge or ``wall" of the truncated
disk. In all these cases, except in CoKu Tau$/$4, material is accreting onto the
star, so gas remains inside the truncated disk, but with a small or negligible amount of 
small dust, making these regions optically thin.  

Here we present models of UX Tau A and LkCa 15, low-mass pre-main sequence stars
in the young, $\sim$1 Myr old Taurus star-forming region which have been
previously reported as transitional disks \citep{furlan06, bergin04}.   
We present evidence for gaps in optically thick disks, as opposed
to ``inner holes", that is, large reductions
of small dust from the star out to an outer optically thick wall.

\section{Observations \& Data Reduction}
\label{sec:obs}
Figures 1 and 2 are plots of the spectral energy distributions (SEDs) for UX Tau
A and LkCa 15 respectively.
For Lk Ca 15, the reduction of the {\it Spitzer} Infrared Array Camera (IRAC)
images (Program 37) was done with SSC pipeline S14.0 and the post-BCD (Basic
Calibrated Data) MOPEX v030106 \citep{makovoz06}. We extracted the photometry of
this object using the apphot package in IRAF, with an aperture radius of 10
pixels and a background annulus from 10 to 20 pixels.  Fluxes at 3.6, 4.5, 5.8,
and 8.0 {\micron} are 7.55, 7.35, 7.24, 6.41 mag (1 ${\sigma}$=0.05 mag) calibrated according to \citet{reach06}. The
Submillimeter Array (SMA) observations of LkCa 15 were made on September 6th,
2003 using the Compact Configuration of six of the 6 meter diameter antennas at
345 GHz with full correlator bandwidth of 2 GHz. Calibration of the visibility
phases and amplitudes was achieved with observations of the quasar 0423-013 and
0530+135, typically at intervals of 20 minutes. Observations of Uranus provided
the absolute scale for the flux density calibration and the uncertainties in the
flux scale are estimated to be 20$\%$. The data were calibrated using the MIR
software package (http://www.cfa.harvard.edu/$\sim$cqi/mircook.html).  Fluxes at
216.5, 226.5, 345.2, and 355.2 GHz are 121.4${\pm}$4.3, 152.8${\pm}$5.2,
416.8${\pm}$37.6, and 453.1${\pm}$48.1 mJy respectively.

\section{Analysis}

\subsection{Model Parameters}
\label{model}

We follow D'Alessio et al. (2006, 2005) to calculate the structure and emission
of the optically thick disk and the wall.    
Input parameters for the optically thick disk are the stellar properties, the
mass accretion rate of the disk ($\mdot$), the viscosity parameter ($\alpha$),
and the settling parameter $\epsilon=\zeta_{up}/\zeta_{st}$, i.e. the mass
fraction of the small grains in the upper layers relative to the standard
dust-to-gas mass ratio \citep{dalessio06}.  We use a grain-size distribution
that follows a power-law of the form $a$$^{-3.5}$, where $a$ is the grain
radius, with a minimum grain size of 0.005 {\micron}.  In the upper, optically
thin layer of the disk, the maximum grain size is 0.25 {\micron}.  In the
midplane of the disk, the maximum grain size is 1 mm.  The radiative transfer in
the wall atmosphere is calculated with the stellar properties, $\mdot$, the
maximum and minimum grain sizes, and the temperature of the optically thin wall
atmosphere (T$_{wall}$).  Table 1 lists the maximum grain sizes of the walls and
other relevant parameters.  We use the same dust composition as
\citet{espaillat07} unless otherwise noted.

In each case we use a distance of 140pc to Taurus \citep{kenyon94}.  We assume
an outer disk radius of 300 AU.  Spectral types and stellar temperatures are
adopted from \citet{KH95}.  Data are dereddened with the \citet{mathis90}
reddening law and extinctions are derived from fitting a standard stellar
photosphere \citep{KH95} to the data.  Stellar parameters (M$_{*}$, R$_{*}$,
L$_{*}$) are derived from the HR diagram and the Baraffe evolutionary tracks
\citep{baraffe02}.  Mass accretion rates are estimated from the U-band excess
following \citet{gullbring98} with a typical uncertainty of a factor of 3
\citep{calvet04}.

\begin{deluxetable}{l c c}
\tabletypesize{\scriptsize}
\tablewidth{0pt}
\tablecaption{Stellar and Model Properties\label{tab:prop}}
\startdata
\hline
\hline
\multicolumn{3}{c}{Stellar Properties}\\
\hline
\hline
\colhead{} & \colhead{UX Tau A}  & \colhead{LkCa 15}\\
\hline
M$_{*}$ (M$_{\sun}$) & 1.5  & 1.1 \\
R$_{*}$ (R$_{\sun}$) & 2  & 1.7 \\
T$_{*}$ (K) & 4900  & 4350 \\
L$_{*}$ ($\lsun$) & 2.18  & .96 \\
$\mdot$ (M$_{\sun}$ yr$^{-1}$) & 9.6$\times$10$^{-9}$ & 2.4$\times$10$^{-9}$\\
Inclination (deg) & 60 & 42$^{1}$\\
A$_{V}$ & 1.3  & 1.2\\
Spectral Type & K2 & K5\\
\hline
\hline
\multicolumn{3}{c}{Optically Thick Inner Wall}\\
\hline
a$_{max}$ ({\micron})$^{2}$ & 10 & 1\\
T$_{wall}$ (K) & 1400  & 1400\\
z$_{wall}$ (AU)$^{2, 3}$ & 0.01  & 0.01\\
R$_{wall}$ (AU) & 0.16 & 0.12\\
\hline
\multicolumn{3}{c}{Optically Thick Inner Disk$^{4}$}\\
\hline
R$_{disk,out}$ (AU)$^{2}$ & $<$0.18  & $<$0.15\\
M$_{disk, inner}$ (M$_{\sun}$) & $<$8$\times$10$^{-6}$ & $<$5$\times$10$^{-5}$\\
\hline
\multicolumn{3}{c}{Optically Thick Outer Wall} \\
\hline
a$_{max}$ ({\micron}) & 0.25  & 0.25\\
T$_{wall}$ (K)$^{2}$ & 110  & 95\\
z$_{wall}$ (AU)$^{2}$ & 6 & 4\\
R$_{wall}$ (AU) & 56 & 46\\
\hline
\multicolumn{3}{c}{Optically Thick Outer Disk}\\
\hline
$\epsilon$$^{2}$ & .01  & .001\\
$\alpha$$^{2}$ & .015  & .0006\\
M$_{disk}$ (M$_{\sun}$)$^{2}$ & .01 & .1\\
\hline
\multicolumn{3}{c}{Optically Thin Inner Region}\\
\hline
R$_{in, thin}$ (AU) & -  & 0.15 (0.12)$^{5}$\\
R$_{out, thin}$ (AU)$^{2}$ & -  & 5 (4) \\
a$_{min, thin}$ ({\micron}) & -  & 0.005\\
a$_{max, thin}$ ({\micron}) & -  & 0.25\\
M$_{dust,thin}$ (M$_{\sun}$) & - & 4(5)$\times$10$^{-11}$
\enddata
\tablenotetext{1}{\citet{simon}}
\tablenotetext{2}{These are free parameters which are constrained by the best
fit to the SED.}
\tablenotetext{3}{z$_{wall}$ is the height of the wall above the midplane}
\tablenotetext{4}{We assume the same $\epsilon$ and $\alpha$ as the outer disk.}
\tablenotetext{5}{For LkCa 15, values in parenthesis refer to parameters in the
case that there is no optically thick inner wall.}
\end{deluxetable}

\begin{figure}
\figurenum{1}
\epsscale{1.2}
\plotone{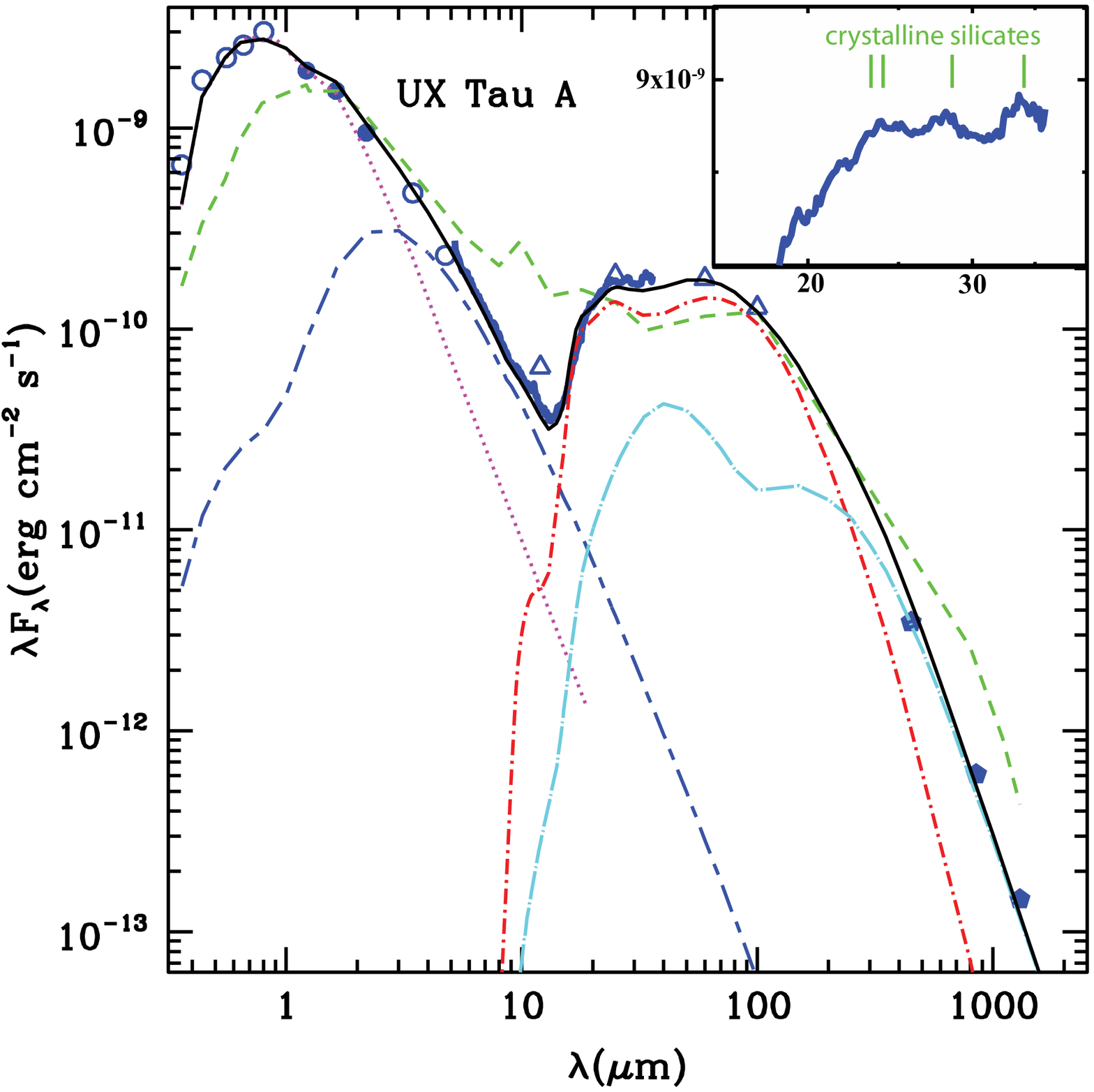}
\caption{SED and model of UX Tau A.  We show ground-based optical, L- and M-band
photometry, \citep[open circles]{KH95}, J,H,K (2MASS; filled circles), 
$Spitzer$ IRS \citep[blue solid line]{furlan06}, IRAS \citep[open
triangles]{weaverjones92}, and millimeter \citep[filled pentagons]{andrews05}
data.  The solid black line is the best fit model with an disk gap of $\sim$56
AU (see Table 1 for model parameters).  Separate model components are as
follows: stellar photosphere (magenta dotted line), inner wall (blue
short-long-dash), outer wall (red dot-short-dash), and outer disk (cyan
dot-long-dash).  We also show the median SED of Taurus (green short-dashed
line).  The insert is a close-up of the {\it Spitzer} IRS spectrum longwards of
$\sim$16 {\micron} and indicates the crystalline silicate emission features in
addition to underlying features from amorphous silicates (Watson et al. 2007,
Sargent et al. in preparation).  [See the electronic edition of the Journal for
a color version of this figure.]
}
\label{figsed}
\end{figure}

\begin{figure}
\figurenum{2}
\epsscale{0.6}
\plotone{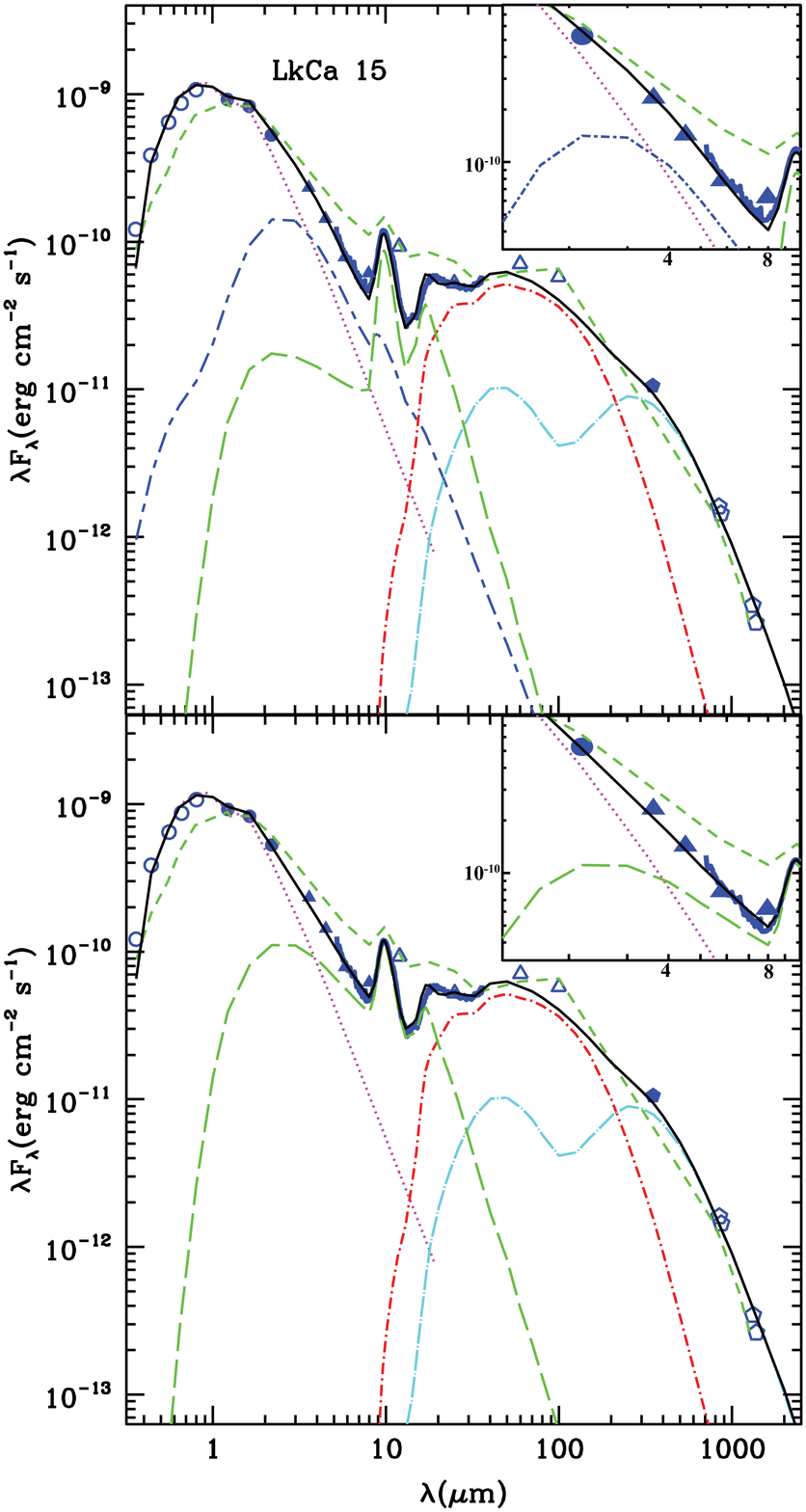}
\caption{SED and model of Lk Ca 15.  Symbols and lines are the same as listed in
Figure 1's caption.  The green long-dash line is the optically thin inner
region.  Open pentagons come from this work.  LkCa 15 has an outer disk that is
truncated at an inner radius of $\sim$46 AU which is consistent with
millimeter results \citep{pietu06}.  Top:  We can fit the SED with an inner optically
thick wall and a small amount of optically thin dust in the gap.  Bottom:  Here
we show a disk model of Lk Ca 15 without an optically thick inner wall.  Model parameters are listed in
Table 1.  The inserts are a close-ups of the SED and models between 3 to
$\sim$10 {\micron}.    [See the electronic edition of the Journal for a color
version of this figure.]
}
\label{figsed}
\end{figure}

\subsection{UX Tau A}

The sharp flux increase in the SED of UX Tau A (Figure 1) can be modeled
by an optically thick outer disk truncated at
$\sim$56 AU. The wall of the outer disk dominates the flux in the
mid- and far-infrared while the outer optically thick disk contributes to most
of the flux in the millimeter. We note that UX Tau A 
is in a multiple system \citep{furlan06}, with the closest component at 2."6 $\sim$ 360 pc
in projection; thus, the outer disk may have a smaller R$_{disk,out}$ than assumed
here, which may decrease the disk mass but not the derived wall properties
a$_{max}$, T$_{wall}$, and z$_{wall}$ which are constrained by the best fit to the mid-infrared.  In modeling the outer disk, we hold $\mdot$ fixed and vary $\alpha$; this is 
equivalent to finding the best-fit disk mass since M$_{d}$ {$\propto$} $\mdot$/$\alpha$.  

When compared to the median SED of Taurus \citep[Fig. 1]{dalessio99, furlan06}
which has been shown to be representative of an optically thick continuous disk
\citep{dalessio99, dalessio06}, UX Tau A's relatively strong mid-infrared
deficit makes it apparent that its disk is not continuous i.e. it is
not a ``full disk".  However, in contrast with the other transitional disks
found around TTS, the near-infrared portion of UX Tau A's SED
agrees with the median SED of TTS in Taurus; this indicates that
optically thick material remains in the innermost part of the disk
in contrast to all other transitional disks modeled so far.
%
%
Since the inner disk is optically thick, it
must have a sharp gas-dust
transition at the dust destruction radius 
as ``full disks" in CTTS
\citep{muzerolle03, dalessio06}.
In Figure 1 we show the
contribution from the wall of an optically thick inner disk located at the dust
destruction radius at 0.16 AU,
assuming T$_{wall}$=1400 K.  
The best fit is obtained with large grains, in agreement with the
lack of a 10 {\micron} silicate feature.
The flux deficit in the SED around 10 {\micron} puts an upper
limit to the extent of the inner optically thick disk to $<$0.18 AU.

\subsection{Lk Ca 15}
\label{lkca15}

Our analysis of a model fit to LkCa 15 shows an optically thick disk truncated
at $\sim$46 AU, as delineated by millimeter interferometric imaging
\citep{pietu06}.  The outer optically thick disk contributes to most of the flux
in the millimeter and the wall of the outer disk contributes much of the flux in
the mid- and far-infrared (Figure 2).

The near-infrared portion of the SED of LkCa 15 lies
below the median SED of Taurus, so the case for an optically thick
inner disk is not as clear cut as in UX Tau A. However, 
the near-IR excess above the photosphere in Lk Ca 15 is substantially
higher than is seen in GM Aur
\citep{calvet05}, TW Hya \citep{calvet02, uchida04}, or CS Cha
\citep{espaillat07}, where the optically thin inner disk contains a small amount of dust.
Our analysis suggests that an optically thick inner disk wall
at the dust destruction radius (0.12 AU) can account for the significant
near-infrared excess in Lk Ca 15 (Fig. 2, top),
and that the optically thick component cannot extend beyond 0.15 AU.
We also require 4$\times$10$^{-11}$ ${\msun}$ of optically thin
dust between 0.15 and 5 AU to produce the 10 {\micron} silicate feature.
This optically thin dust mixture is composed of 85$\%$ amorphous silicates,
6.8$\%$ organics, 1.3$\%$ troilite, 6.8$\%$ amorphous carbon, and less than
1$\%$ enstatite and forsterite.  The total emission of this optically thin
region is scaled to the vertical optical depth at 10 {\micron}, $\tau_{0}$
$\sim$ 0.012.  We follow \citep{calvet02} in calculating the optically thin dust
region and note that the dust composition and $\tau_{0}$ are free parameters
constrained by the best fit to the SED.    

In the bottom panel of Figure 2 we present an alternative structure for LkCa
15's inner disk.   It is possible to fit LkCa 15's SED with only optically thin
dust within the inner hole.  In this model, the near-infrared excess and 10
{\micron} emission would originate in 5$\times$10$^{-11}$ M$_{\sun}$ of
optically thin dust located between 0.12 to 4 AU.  This optically thin dust
mixture would be composed of 61$\%$ amorphous silicates, 7$\%$ organics, 1$\%$
troilite, 30$\%$ amorphous carbon, and less than 1$\%$ enstatite and forsterite.
The total emission of this optically thin region is scaled to the vertical
optical depth at 10 {\micron}, $\tau_{0} \sim 0.018$.  This model (bottom inset,
Fig. 2) does not fit the slope of the near-side of the IRS spectrum
($<$7{\micron}) or the IRAC data 
as well as the previously discussed model (top inset, Fig. 2).  The
optical depth in the near-infrared is 
$\sim$0.01, about 5 times greater than that of the
previous model, mainly due to the difference in the amorphous carbon fraction.  

While an optically thin region is necessary in both scenarios, we can exclude a
model where the optically thin region extends further than $\sim$ 5 AU since the
contribution at 20 {\micron} then becomes too strong.

\section{Discussion \& Conclusions}

Here we introduce the ``pre-transitional disk" class where we see the 
incipient
development
of disk gaps in 
optically thick 
protoplanetary disks as evidenced by significant near-infrared
excesses when compared to the Taurus median SED and previously studied
transitional disks \citep{dalessio05, calvet02, uchida04, calvet05,
espaillat07}.  
The pre-transitional disk of UX Tau A has a $\sim$56 AU gap as
opposed to an inner hole.  It is also possible to fit LkCa 15's SED with a
$\sim$46 AU gap that contains some optically thin dust; a model that has a hole
rather than a gap also fits its SED and future near-infrared interferometry may
be able to discriminate between these models.  However, the truncation of LkCa
15's outer disk at $\sim$46 AU is consistent with resolved millimeter
interferometric observations \citep{pietu06} which makes it one of three inner
disk holes imaged in the millimeter (TW Hya: Hughes et al. 2007; GM Aur: Wilner
et al. in preparation).  In addition to our sample, the disks around F-G stars
studied by \citet{brown07} also belong to the pre-transitional disk category.  The large gaps that are being detected in pre-transitional disks are most likely due to observational bias since larger gaps will create larger mid-infrared deficits in the SED.  Smaller gaps will most likely have less apparent dips in their SEDS and be more difficult to identify, however, if their gaps contain some optically thin material the silicate emission in these objects should be much stronger than can be explained by a full disk model.  

The existence of an inner optically thick disk may be an indicator of the first
stages of disk clearing that will eventually lead to the the inner holes that
have been seen in previously reported transitional disks; this has important
implications on disk evolution theories since only planet-formation can account
for this structure. 
Hydrodynamical simulations have shown that a newly formed planet could accrete
and sweep out the material around it through tidal disturbances and this is
sufficient in producing the hole size in CoKu Tau$/$4 \citep{quillen04}, even
maintaining substantial accretion rates \citep{varniere06}.  
Moreover, \citet{najita07} have found that the intrinsic properties
of transitional disks may favor planet formation.
Another proposed formation mechanism for the holes in transitional disks
is photoevaporation, in which a photoevaporative wind halts
mass accretion towards the inner disk
and material in this inner disk is rapidly evacuated creating
an inner hole \citep{clarke01}; the hole then increases in size as the edge
continues
photoevaporating \citep{alexanderarmitage}.
Neither this model nor the inside-out evacuation induced by the MRI
\citep{chiang07} would
explain how an optically thick disk accreting at a sizable
accretion rate (see Table 1) would remain inside the hole.
Rapid dust growth and settling has also been proposed to explain the holes in
disks \citep{lin04}.  Again, this does not account for the presence of optically
thick inner disk material given that theory suggests grain growth should be
fastest in the inner disk, not at some intermediate radius \citep{weiden97,
chiang99}.
      
Our sample also has interesting dust compositions (Watson et al. 2007, Sargent
et al. in preparation).  LkCa 15 has an amorphous silicate feature indicating
little if any processing leading to the crystallization seen in other young
stars.  Amorphous silicates are also seen in CoKu Tau$/4$, DM Tau, and GM Aur. 
In contrast, UX Tau A is different from all the other transitional disks because
it has crystalline silicate emission features in addition to amorphous silicate
emission features (Fig. 1 inset).  
The wall at $\sim$56 AU is the main contributor to the crystalline silicate
emission since it dominates the flux in the mid- and far-infrared.  This raises
the question of whether crystalline silicates are created close to the star or
if they can be created in situ at $\sim$56 AU.  If the former, it challenges
current radial-mixing theories, none of which can get significant amounts of
crystalline silicates out to this distance \citep{gail01, kellergail04}.  One
possibility for {\it in situ} processing may be
collisions of larger bodies, which might produce small grains heated
sufficiently to create crystals (S. Kenyon, personal communication).  

Pre-transitional disks offer further insight into the diversity of the
``transitional disk" class and future studies of these disks will greatly
advance our understanding of disk evolution and planet formation.

\vskip -0.1in \acknowledgments{We thank the anonymous referee, E. Bergin, W.
Forrest, S. Kenyon, K. H. Kim, J. Miller, B. Sargent, and M. Zhao for insightful
discussions.  This work is based on observations made with the Spitzer Space
Telescope.  N.C. and L.H. acknowledge support from NASA
Origins Grants NNG05GI26G and NNG06GJ32G.  P.D. acknowledges grants from
CONACyT, M\`exico. D.M.W. acknowledges support from the Spitzer Infrared
Spectrograph
Instrument Project.}

\end{document}